\documentclass[sigconf,9pt,nonacm]{acmart}

\AtBeginDocument{%
  }

\usepackage{balance}
\usepackage[countmax]{subfloat}
\usepackage{subcaption}
\usepackage{amsmath,amsfonts}
\usepackage{algorithmic}
\usepackage{graphicx}
\usepackage{textcomp}
\usepackage{xcolor}
\usepackage{lineno}
\DeclareGraphicsExtensions{.png,.pdf,.jpg}
\graphicspath{{img/} {./}}

\usepackage[nolist,nohyperlinks,withpage]{acronym}
\usepackage[normalem]{ulem}
\usepackage{enumitem}

\acrodef{ADC}[ADC]{Analog to Digital Converter}
\acrodef{ADEXP}[AdExp-I\&F]{Adaptive-Exponential Integrate and Fire}
\acrodef{AER}[AER]{Address-Event Representation}
\acrodef{AEX}[AEX]{AER EXtension board}
\acrodef{AE}[AE]{Address-Event}
\acrodef{AFM}[AFM]{Atomic Force Microscope}
\acrodef{AGC}[AGC]{Automatic Gain Control}
\acrodef{AI}[AI]{Artificial Intelligence}
\acrodef{AMDA}[AMDA]{AER Motherboard with D/A converters}
\acrodef{ANN}[ANN]{Artificial Neural Network}
\acrodef{API}[API]{Application Programming Interface}
\acrodef{APMOM}[APMOM]{Alternate Polarity Metal On Metal}
\acrodef{ARM}[ARM]{Advanced RISC Machine}
\acrodef{ASIC}[ASIC]{Application Specific Integrated Circuit}
\acrodef{AdExp}[AdExp-IF]{Adaptive Exponential Integrate-and-Fire}
\acrodef{BCM}[BMC]{Bienenstock-Cooper-Munro}
\acrodef{BD}[BD]{Bundled Data}
\acrodef{BEOL}[BEOL]{Back-end of Line}
\acrodef{BG}[BG]{Bias Generator}
\acrodef{BMI}[BMI]{Brain-Machine Interface}
\acrodef{BTB}[BTB]{band-to-band tunnelling}
\acrodef{CAD}[CAD]{Computer Aided Design}
\acrodef{CAM}[CAM]{Content Addressable Memory}
\acrodef{CAVIAR}[CAVIAR]{Convolution AER Vision Architecture for Real-Time}
\acrodef{CA}[CA]{Cortical Automaton}
\acrodef{CCN}[CCN]{Cooperative and Competitive Network}
\acrodef{CDR}[CDR]{Clock-Data Recovery}
\acrodef{CFC}[CFC]{Current to Frequency Converter}
\acrodef{CHP}[CHP]{Communicating Hardware Processes}
\acrodef{CMIM}[CMIM]{Metal-insulator-metal Capacitor}
\acrodef{CML}[CML]{Current Mode Logic}
\acrodef{CMOL}[CMOL]{Hybrid CMOS nanoelectronic circuits}
\acrodef{CMOS}[CMOS]{Complementary Metal-Oxide-Semiconductor}
\acrodef{CNN}[CCN]{Convolutional Neural Network}
\acrodef{COTS}[COTS]{Commercial Off-The-Shelf}
\acrodef{CPG}[CPG]{Central Pattern Generator}
\acrodef{CPLD}[CPLD]{Complex Programmable Logic Device}
\acrodef{CPU}[CPU]{Central Processing Unit}
\acrodef{CSM}[CSM]{Cortical State Machine}
\acrodef{CSP}[CSP]{Constraint Satisfaction Problem}
\acrodef{CTXCTL}[CTXCTL]{CortexControl}
\acrodef{CV}[CV]{Coefficient of Variation}
\acrodef{DAC}[DAC]{Digital to Analog Converter}
\acrodef{DAS}[DAS]{Dynamic Auditory Sensor}
\acrodef{DAVIS}[DAVIS]{Dynamic and Active Pixel Vision Sensor}
\acrodef{DBN}[DBN]{Deep Belief Network}
\acrodef{DFA}[DFA]{Deterministic Finite Automaton}
\acrodef{DIBL}[DIBL]{drain-induced-barrier-lowering}
\acrodef{DI}[DI]{delay insensitive}
\acrodef{DMA}[DMA]{Direct Memory Access}
\acrodef{DNF}[DNF]{Dynamic Neural Field}
\acrodef{DNN}[DNN]{Deep Neural Network}
\acrodef{DOF}[DOF]{Degrees of Freedom}
\acrodef{DPE}[DPE]{Dynamic Parameter Estimation}
\acrodef{DPI}[DPI]{Differential Pair Integrator}
\acrodef{DRAM}[DRAM]{Dynamic Random Access Memory}
\acrodef{DRRZ}[DR-RZ]{Dual-Rail Return-to-Zero}
\acrodef{DR}[DR]{Dual Rail}
\acrodef{DSP}[DSP]{Digital Signal Processor}
\acrodef{DVS}[DVS]{Dynamic Vision Sensor}
\acrodef{DYNAP}[DYNAP]{Dynamic Neuromorphic Asynchronous Processor}
\acrodef{EBL}[EBL]{Electron Beam Lithography}
\acrodef{EDVAC}[EDVAC]{Electronic Discrete Variable Automatic Computer}
\acrodef{EEG}[EEG]{electroencephalography}
\acrodef{EIN}[EIN]{Excitatory-Inhibitory Network}
\acrodef{EM}[EM]{Expectation Maximization}
\acrodef{EMG}[EMG]{electromygraphy}
\acrodef{EPSC}[EPSC]{Excitatory Post-Synaptic Current}
\acrodef{EPSP}[EPSP]{Excitatory Post-Synaptic Potential}
\acrodef{EZ}[EZ]{Epileptogenic Zone}
\acrodef{FDSOI}[FDSOI]{Fully-Depleted Silicon on Insulator}
\acrodef{FET}[FET]{Field-Effect Transistor}
\acrodef{FFT}[FFT]{Fast Fourier Transform}
\acrodef{FI}[F-I]{Frequency-Current}
\acrodef{FPGA}[FPGA]{Field Programmable Gate Array}
\acrodef{FR}[FR]{Fast Ripple}
\acrodef{FSA}[FSA]{Finite State Automaton}
\acrodef{FSM}[FSM]{Finite State Machine}
\acrodef{FVF}[FVF]{Flipped Voltage Follower}
\acrodef{GIDL}[GIDL]{gate-induced-drain-leakage}
\acrodef{GOPS}[GOPS]{Giga-Operations per Second}
\acrodef{GPU}[GPU]{Graphical Processing Unit}
\acrodef{GUI}[GUI]{Graphical User Interface}
\acrodef{HAL}[HAL]{Hardware Abstraction Layer}
\acrodef{HFO}[HFO]{High Frequency Oscillation}
\acrodef{HH}[H\&H]{Hodgkin \& Huxley}
\acrodef{HMM}[HMM]{Hidden Markov Model}
\acrodef{HRS}[HRS]{High-Resistive State}
\acrodef{HR}[HR]{Human Readable}
\acrodef{HSE}[HSE]{Handshaking Expansion}
\acrodef{HW}[HW]{Hardware}
\acrodef{ICT}[ICT]{Information and Communication Technology}
\acrodef{IC}[IC]{Integrated Circuit}
\acrodef{iEEG}[iEEG]{intracranial electroencephalography}
\acrodef{IF2DWTA}[IF2DWTA]{Integrate \& Fire 2--Dimensional WTA}
\acrodef{IFSLWTA}[IFSLWTA]{Integrate \& Fire Stop Learning WTA}
\acrodef{IF}[I\&F]{Integrate-and-Fire}
\acrodef{IMU}[IMU]{Inertial Measurement Unit}
\acrodef{INCF}[INCF]{International Neuroinformatics Coordinating Facility}
\acrodef{INI}[INI]{Institute of Neuroinformatics}
\acrodef{IO}[I/O]{Input/Output}
\acrodef{IPSC}[IPSC]{Inhibitory Post-Synaptic Current}
\acrodef{IPSP}[IPSP]{Inhibitory Post-Synaptic Potential}
\acrodef{IP}[IP]{Intellectual Property}
\acrodef{ISI}[ISI]{Inter-Spike Interval}
\acrodef{IoT}[IoT]{Internet of Things}
\acrodef{JFLAP}[JFLAP]{Java - Formal Languages and Automata Package}
\acrodef{LEDR}[LEDR]{Level-Encoded Dual-Rail}
\acrodef{LFP}[LFP]{Local Field Potential}
\acrodef{LLC}[LLC]{Low Leakage Cell}
\acrodef{LNA}[LNA]{Low-Noise Amplifier}
\acrodef{LPF}[LPF]{Low Pass Filter}
\acrodef{LRS}[LRS]{Low-Resistive State}
\acrodef{LSM}[LSM]{Liquid State Machine}
\acrodef{LTD}[LTD]{Long Term Depression}
\acrodef{LTI}[LTI]{Linear Time-Invariant}
\acrodef{LTP}[LTP]{Long Term Potentiation}
\acrodef{LTU}[LTU]{Linear Threshold Unit}
\acrodef{LUT}[LUT]{Look-Up Table}
\acrodef{LIF}[LIF]{Leaky Integrate and Fire}
\acrodef{LVDS}[LVDS]{Low Voltage Differential Signaling}
\acrodef{MCMC}[MCMC]{Markov-Chain Monte Carlo}
\acrodef{MCU}[MCU]{Microcontroller Unit}
\acrodef{MEMS}[MEMS]{Micro Electro Mechanical System}
\acrodef{MFR}[MFR]{Mean Firing Rate}
\acrodef{MIM}[MIM]{Metal Insulator Metal}
\acrodef{MLP}[MLP]{Multilayer Perceptron}
\acrodef{MOSCAP}[MOSCAP]{Metal Oxide Semiconductor Capacitor}
\acrodef{MOSFET}[MOSFET]{Metal Oxide Semiconductor Field-Effect Transistor}
\acrodef{MOS}[MOS]{Metal Oxide Semiconductor}
\acrodef{MRI}[MRI]{Magnetic Resonance Imaging}
\acrodef{NDFSM}[NDFSM]{Non-deterministic Finite State Machine}
\acrodef{ND}[ND]{Noise-Driven}
\acrodef{NEF}[NEF]{Neural Engineering Framework}
\acrodef{NHML}[NHML]{Neuromorphic Hardware Mark-up Language}
\acrodef{NIL}[NIL]{Nano-Imprint Lithography}
\acrodef{NMDA}[NMDA]{N-Methyl-D-Aspartate}
\acrodef{NME}[NE]{Neuromorphic Engineering}
\acrodef{NN}[NN]{Neural Network}
\acrodef{NRZ}[NRZ]{Non-Return-to-Zero}
\acrodef{NSM}[NSM]{Neural State Machine}
\acrodef{OR}[OR]{Operating Room}
\acrodef{OTA}[OTA]{Operational Transconductance Amplifier}
\acrodef{PCB}[PCB]{Printed Circuit Board}
\acrodef{PCHB}[PCHB]{Pre-Charge Half-Buffer}
\acrodef{PCM}[PCM]{Phase Change Memory}
\acrodef{PE}[PE]{Phase Encoding}
\acrodef{PFA}[PFA]{Probabilistic Finite Automaton}
\acrodef{PFC}[PFC]{prefrontal cortex}
\acrodef{PFM}[PFM]{Pulse Frequency Modulation}
\acrodef{PR}[PR]{Production Rule}
\acrodef{PSC}[PSC]{Post-Synaptic Current}
\acrodef{PSP}[PSP]{Post-Synaptic Potential}
\acrodef{PSTH}[PSTH]{Peri-Stimulus Time Histogram}
\acrodef{QDI}[QDI]{Quasi Delay Insensitive}
\acrodef{RAM}[RAM]{Random Access Memory}
\acrodef{RA}[RA]{Resected Area}
\acrodef{RDF}[RDF]{random dopant fluctuation}
\acrodef{RELU}[ReLu]{Rectified Linear Unit}
\acrodef{RLS}[RLS]{Recursive Least-Squares}
\acrodef{RMSE}[RMSE]{Root Mean Squared-Error}
\acrodef{RMS}[RMS]{Root Mean Squared}
\acrodef{RNN}[RNN]{Recurrent Neural Networks}
\acrodef{ROLLS}[ROLLS]{Reconfigurable On-Line Learning Spiking}
\acrodef{RRAM}[R-RAM]{Resistive Random Access Memory}
\acrodef{R}[R]{Ripples}
\acrodef{SAC}[SAC]{Selective Attention Chip}
\acrodef{SAT}[SAT]{Boolean Satisfiability Problem}
\acrodef{SCX}[SCX]{Silicon CorteX}
\acrodef{SD}[SD]{Signal-Driven}
\acrodef{SEM}[SEM]{Spike-based Expectation Maximization}
\acrodef{SLAM}[SLAM]{Simultaneous Localization and Mapping}
\acrodef{SNN}[SNN]{Spiking Neural Network}
\acrodef{SNR}[SNR]{Signal to Noise Ratio}
\acrodef{SOC}[SOC]{System-On-Chip}
\acrodef{SOI}[SOI]{Silicon on Insulator}
\acrodef{SOZ}[SOZ]{Seizure Onset Zone}
\acrodef{SP}[SP]{Separation Property}
\acrodef{SRAM}[SRAM]{Static Random Access Memory}
\acrodef{STDP}[STDP]{Spike-Timing Dependent Plasticity}
\acrodef{STD}[STD]{Short-Term Depression}
\acrodef{STP}[STP]{Short-Term Plasticity}
\acrodef{STT-MRAM}[STT-MRAM]{Spin-Transfer Torque Magnetic Random Access Memory}
\acrodef{STT}[STT]{Spin-Transfer Torque}
\acrodef{SW}[SW]{Software}
\acrodef{TCAM}[TCAM]{Ternary Content-Addressable Memory}
\acrodef{TFT}[TFT]{Thin Film Transistor}
\acrodef{TLE}[TLE]{Temporal Lobe Epilepsy}
\acrodef{USB}[USB]{Universal Serial Bus}
\acrodef{VHDL}[VHDL]{VHSIC Hardware Description Language}
\acrodef{VLSI}[VLSI]{Very Large Scale Integration}
\acrodef{VOR}[VOR]{Vestibulo-Ocular Reflex}
\acrodef{WCST}[WCST]{Wisconsin Card Sorting Test}
\acrodef{WTA}[WTA]{Winner-Take-All}
\acrodef{XML}[XML]{eXtensible Mark-up Language}
\acrodef{divmod3}[DIVMOD3]{divisibility of a number by three}
\acrodef{hWTA}[hWTA]{hard Winner-Take-All}
\acrodef{sWTA}[sWTA]{soft Winner-Take-All}
\acrodef{EEG}[EEG]{Electroencephalography}
\acrodef{iEEG}[iEEG]{Intracranial EEG}
\acrodef{ECoG}[ECoG]{Electrocorticography}
\acrodef{O2G}[O2G]{ODE to GDS}
\acrodef{HEiL}[HEiL]{Hardware-Emulation-in-the-loop}
\acrodef{NSF}[NSF]{National Science Foundation}
\acrodef{CISE}[CISE]{Computer and Information Science and Engineering}
\acrodef{FutureCoRe}[Future CoRe]{Future Computing Research}
\acrodef{SHF}[SHF]{Systems and Hardware Foundations}
\acrodef{RI}[RI]{Robust Intelligence}
\acrodef{FET}[FET]{Future of Work at the Human-Technology Frontier} 
\acrodef{CSR}[CSR]{Cyber-Physical Systems and Smart and Connected Communities} 
\acrodef{CER}[CER]{Computing Education Research}
\acrodef{O2G}[O2G]{ODE-to-GDS}
\acrodef{EDA}[EDA]{Electronic Design Automation}
\acrodef{IR}[IR]{Intermediate Representation}
\acrodef{API}[API]{Application Programming Interface}
\acrodef{FoM}[FoM]{Figure-of-Merit}
\acrodef{MVD}[MVD]{Minimum Viable Demonstrator}
\acrodef{IP}[IP]{Intellectual Property}
\acrodef{IC}[IC]{Integrated Circuit}
\acrodef{SoC}[SoC]{System-on-Chip}
\acrodef{MPW}[MPW]{Multi-Project Wafer}
\acrodef{PDK}[PDK]{Process Design Kit}
\acrodef{DRC}[DRC]{Design Rule Check}
\acrodef{SPICE}[SPICE]{Simulation Program with Integrated Circuit Emphasis}
\acrodef{LVS}[LVS]{Layout Versus Schematic}
\acrodef{PEX}[PEX]{Parasitic Extraction}
\acrodef{PVT}[PVT]{Process, Voltage, and Temperature}
\acrodef{SNR}[SNR]{Signal-to-Noise Ratio}
\acrodef{ADC}[ADC]{Analog-to-Digital Converter}
\acrodef{DAC}[DAC]{Digital-to-Analog Converter}
\acrodef{RF}[RF]{Radio Frequency}
\acrodef{PV}[PV]{Process Variations}
\acrodef{RNN}[RNN]{Recurrent Neural Network}
\acrodef{CRN}[CRN]{Critical Recurrent Network}
\acrodef{CtD}[CtD]{computation-through-dynamics}
\acrodef{STDP}[STDP]{Spike-Timing-Dependent Plasticity}
\acrodef{ODE}[ODE]{Ordinary Differential Equation}
\acrodef{SDE}[SDE]{Stochastic Differential Equation}
\acrodef{FTLE}[FTLE]{Finite-Time Lyapunov Exponent}
\acrodef{LLE}[LLE]{Largest Lyapunov Exponent}
\acrodef{Lyap}[Lyap]{Lyapunov}
\acrodef{ML}[ML]{Machine Learning}
\acrodef{AI}[AI]{Artificial Intelligence}
\acrodef{DL}[DL]{Deep Learning}
\acrodef{RL}[RL]{Reinforcement Learning}
\acrodef{RLHF}[RLHF]{Reinforcement Learning with Human Feedback}
\acrodef{GA}[GA]{Genetic Algorithm}
\acrodef{LHS}[LHS]{Latin hypercube sampling}
\acrodef{PINN}[PINN]{Physics Inspired Neural Network}
\acrodef{BO}[BO]{Bayesian Optimization}
\acrodef{GP}[GP]{Gaussian Process}
\acrodef{MCMC}[MCMC]{Markov Chain Monte Carlo}
\acrodef{MC}[MC]{Monte Carlo}
\acrodef{HEiL}[HEiL]{hardware-emulation-in-the-loop}
\acrodef{KPI}[KPI]{Key Performance Indicator}
\acrodef{QoR}[QoR]{Quality of Results}
\acrodef{DRI}[DRI]{Design Representation Index}
\acrodef{HAL}[HAL]{Hardware Adaptation Latency}
\acrodef{EDR}[EDR]{Energy--Dynamics Ratio}
\acrodef{CPU}[CPU]{Central Processing Unit}
\acrodef{GPU}[GPU]{Graphics Processing Unit}
\acrodef{TPU}[TPU]{Tensor Processing Unit}
\acrodef{ASIC}[ASIC]{Application-Specific Integrated Circuit}
\acrodef{CIM}[CIM]{Compute-in-Memory}
\acrodef{IoIBS}[IoIBS]{Intranet of In-Body Sensors}
\acrodef{NESTml}[NESTml]{NEural Simulation Tool modelling language}
\acrodef{AFE}[AFE]{Analog Front-End}
\acrodef{PFM}[PFM]{Pulse Frequency Modulation}
\acrodef{ADM}[ADM]{Asynchronous Delta Modulator}
\acrodef{aADM}[aADM]{adaptive Asynchronous Delta Modulator}
\acrodef{ASC}[ASC]{Analog to Spike Converter}
\acrodef{PGA}[PGA]{Programmable Gain Amplifier}
\acrodef{FVF}[FVF]{Flipped Voltage Follower}
\acrodef{BPF}[BPF]{Band-Pass Filter}
\acrodef{CDAC}[CDAC]{Capacitor DAC}
\acrodef{VDAC}[VDAC]{Voltage DAC}
\acrodef{SPI}[SPI]{Serial Peripheral Interface}
\acrodef{OTA}[OTA]{Operational Transconductance Amplifier}
\acrodef{SF}[SF]{Source Follower}
\acrodef{IF}[IF]{Intregrate and Fire}
\acrodef{PFM}[PFM]{Pulse Frequency Modulation}
\acrodef{BCI}[BCI]{Brain Computer Interface}

\usepackage{xcolor}
\newcommand{\citehere}[1]{\textcolor{red}{(cite here)}}

\begin{document}

\title{A 32-channel event-based bio-signal analog front-end with adaptive delta and pulse frequency encoding}
\author{Narayanan Shyam, Saptarshi Ghosh, and Giacomo Indiveri}
\affiliation{
  \institution{Institute of Neuroinformatics, University of Zurich and ETH Zurich}
  \city{Zurich}
  \country{Switzerland}
  }

\begin{abstract}
Low-power event-based \acp{AFE} are essential for building efficient, end-to-end neuromorphic signal processing systems.
In this paper, we present an event-based \ac{AFE} \ac{ASIC} optimized for biomedical signal acquisition and encoding.
The chip features 32 independently programmable input channels with dual-mode encoding mechanism outputs, comprising \ac{PFM} and \ac{aADM} circuits.
The \ac{aADM} encoder provides an auto-scaling mechanism that adapts the encoding data-rate  based on the input signal envelope in real-time, enabling very high data compression for low-power information transmission.
This approach paves the way toward adaptive wireless communication of neural signals for on-line processing in brain-computer interfaces.
Fabricated in a 180\,nm CMOS process, the proposed \ac{ASIC} offers a highly configurable interface compatible with state-of-the-art \ac{SNN} neuromorphic processors.
\end{abstract}


\maketitle
\section{Introduction}
\acfp{SNN} implemented on bio-inspired neuromorphic hardware provide a low-power, event-driven signal processing paradigm, ideally suited for handling streaming data on the edge~\cite{Corradi_etal25,Mohan_etal25,Indiveri25}.
To leverage \ac{SNN}-based data processing pipelines effectively, highly efficient event-based front-ends are required to convert analog sensory signals into the appropriate \ac{AER}~\cite{guo2021neural}.

Several bio-signal processing systems have already been proposed in the past that included dedicated asynchronous circuits for low-power and low-latency event-based encoding~\cite{Li_etal13a,Judy_etal14,Corradi_Indiveri15,Van-Assche_Gielen20,Narayanan_etal23,ke20251024}.
However, typically those encoding circuits use fixed thresholds for data conversion.
This strategy faces severe limitations when scaling the converters to large multi-channels or operating them long-term in real world \ac{BCI} scenarios~\cite{Yasar_etal24,Bartels_etal25}: fixing the thresholds too closely will generate large amounts of events for noisy signals; conversely, placing the thresholds too far apart to avoid encoding noise fluctuations will also reduce the encoding accuracy of the signal, degrading the downstream feature extraction.
To overcome these limitations, we present the first realization of an \ac{aADM} able to tune the encoding to the noise levels present in the bio-medical signal.
Specifically, we designed and fabricated a
32-channel \ac{ASIC} with configurable and adaptive sensory signal conditioning for efficient event encoding.
The \ac{AFE} circuits are explicitly optimized to operate at the low end of the frequency spectrum to process biomedical signals.
The novel implementation of the \ac{aADM} circuit dynamically adjusts the delta-modulation threshold in real-time, by following the envelope of the input signal.
This allows it to encode only the meaningful parts of the signal, and to reject the noise floor present in the signal.
By configuring the amount of adaptation, this circuit can adjust the data encoding and  compression ratio at the sensor level, providing a trade-off between desired raw signal reconstruction accuracy and on-line information processing needs (e.g., to just detect the occurrence of a real action potential, or to classify specific spatio-temporal patterns).
This approach makes this \ac{ASIC} suitable for both scaling up to massive multi-channel bio-signal processing and encoding arrays, and for transmitting multi-channel asynchronous \ac{AER} events with low-bandwidth and low-latency.

\section{ASIC architecture}
\begin{figure}
  \centering
  \includegraphics[width=0.8\linewidth]{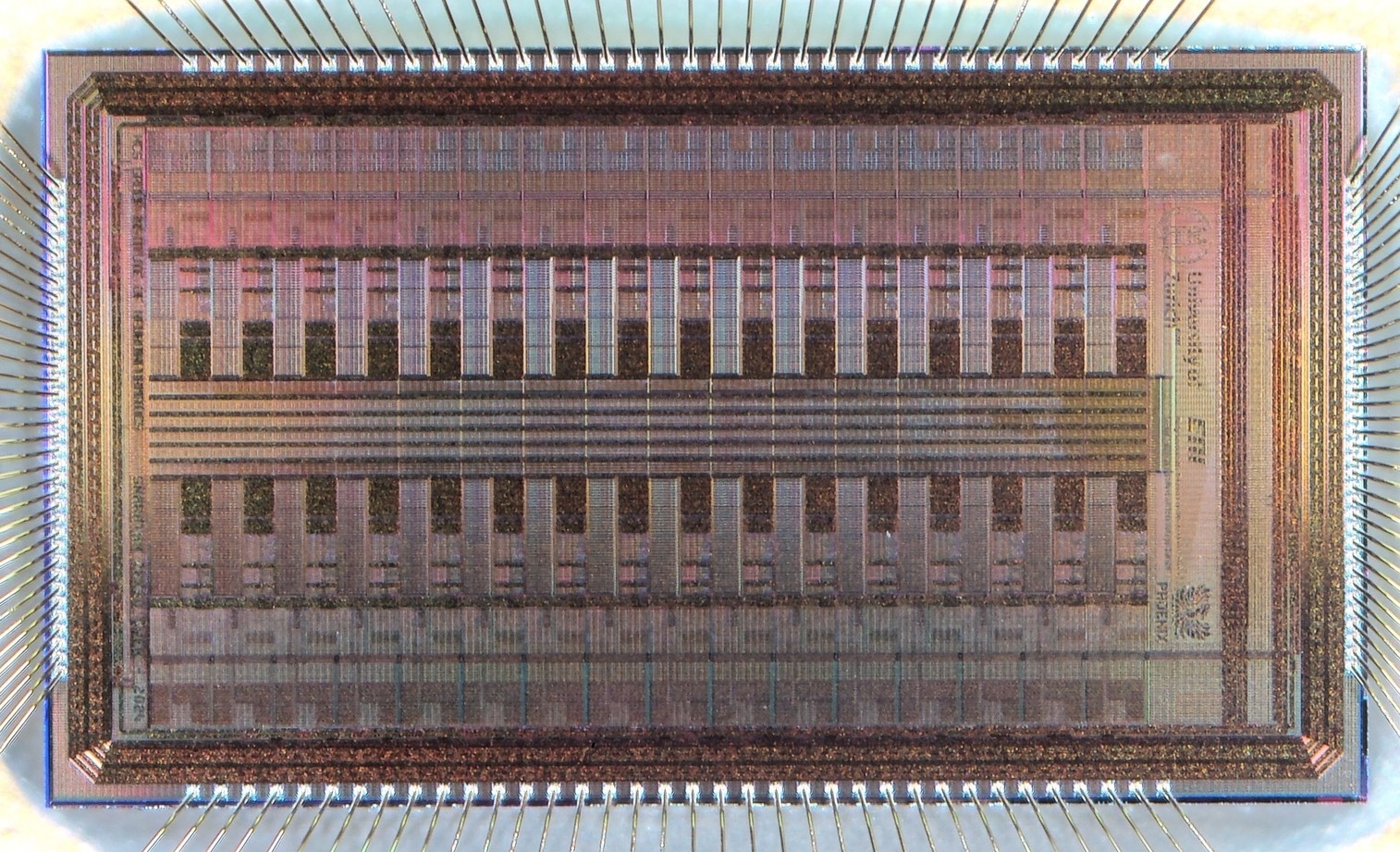}
  \Description{\ac{AFE} chip micrograph, with 32 mixed-signal analog/digital processing blocks, a synchronous digital \ac{SPI} input interface, and an asynchronous \ac{AER} output interface.}
  \caption{\ac{AFE} chip micrograph, with 32 mixed-signal analog/digital processing blocks, a synchronous digital \ac{SPI} input interface, and an asynchronous \ac{AER} output interface.}
  \label{fig:chip}
\end{figure}

The chip comprises 32 analog processing channels that pre-condition the signal with amplification, filtering stages, and event-encoding circuits that convert the input signal into asynchronous discrete events, using parallel encoding methods: pulse-frequency modulation, via a low-power \ac{LIF} neuron, and adaptive delta-modulation, via an \ac{aADM} circuit.
The analog circuit parameters can be programmed using an on-chip temperature compensated bias generator~\cite{Delbruck_Van-Schaik05,Delbruck_etal10}.
All chip configurations, including the bias-generator, an 8\,bit \ac{CDAC} for filter parameters, and a 7\,bit \ac{VDAC} for \ac{aADM} block controls, 
are programmed via a digital \ac{SPI} based interface.
The asynchronous output events generated by each channel are channeled through an arbiter tree and transmitted off-chip using the \ac{AER} communication protocol~\cite{Deiss_etal94,Boahen00}.
The \ac{ASIC} measures $3.22\,\text{mm} \times 5.67\,\text{mm}$ and is fabricated in the standard XFAB 180\,nm.
A micrograph of the fabricated chip is shown in Fig.~\ref{fig:chip}.

\begin{figure*}
  \centering
  \includegraphics[width= 0.95 \linewidth]{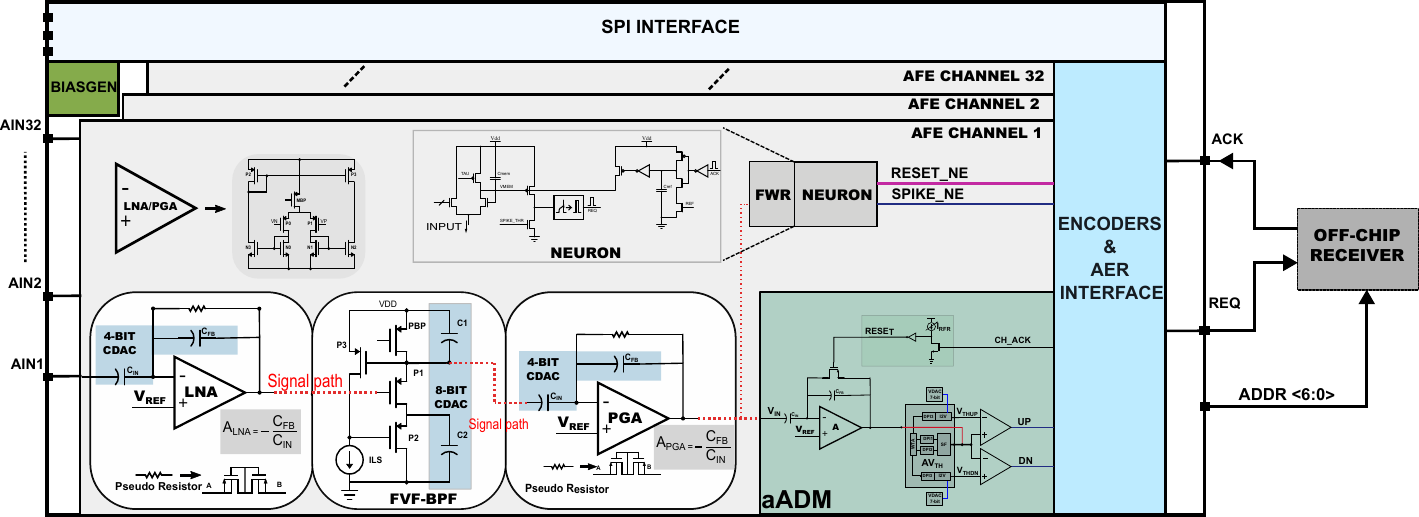}
  \Description{Chip block diagram detailing one of the 32 analog channels, SPI, and AER interface blocks.}
  \caption{Chip block diagram detailing one of the 32 analog channels connected to the shared BiasGen, SPI, and AER interface blocks.}
  \label{fig:architecture}
\end{figure*}

An illustration of the \ac{ASIC} building blocks and its signal processing pipeline is presented in Fig.~\ref{fig:architecture}.
The processing chain for each input channel consists of four main stages: a \ac{LNA}, a fourth-order \ac{BPF}, a \ac{PGA}, and the dual-mode event-based encoding block (\ac{aADM} and \ac{PFM}).

\subsection{Pre-encoding circuit implementation}
\label{sec:pre-encoding-circuit}

\subsubsection{Low noise gain stages (\ac{LNA} and \ac{PGA}):}
For modularity, both the LNA and PGA rely on the same \ac{OTA} core, which is based on a wide input range current-mirror-type architecture to accommodate large input variations~\cite{Harrison_Charles03}.
The amplifiers are operated in a closed loop as capacitive feedback amplifiers utilizing a DC-Servo loop implemented with pseudo-resistors~\cite{Livanelioglu_etal22,Narayanan_etal23}.
The \ac{LNA} amplifies weak input signals with a tunable gain ranging from 0\,dB to 22\,dB, controlled by a 4\,bit binary weighted \ac{CDAC}.

\subsubsection{The \acf{BPF}}
Every \ac{AFE} channel has a 4th-order \ac{BPF} with tunable center frequency $\omega_{0}$ and $Q$.
The filter is based on a \ac{FVF} topology (see Fig.~\ref{fig:architecture}) to achieve better noise performance due to the inherent current reuse present in the architecture~\cite{Kim_Liu23}.
The filter's center frequency $\omega_{0}$ is tunable with the on-chip bias generator~\cite{Delbruck_Van-Schaik05}.
The capacitors of this filter are implemented as an 8-bit \ac{CDAC}.
By configuring $C_1$ and $C_2$, $Q$ and center frequency can be adjusted in a precise way with a resolution of $C_{1,2}/256$ steps, as described in Eq.~(\ref{eq:omegaq}).
\begin{equation}
    \omega_0 = \sqrt \frac { gm_1 \cdot gm_2}{C_1\cdot C_2}, \\
    Q = \sqrt \frac { gm_2 \cdot C_2}{gm_1\cdot C_1}
    \label{eq:omegaq}
  \end{equation}

As each filter's center frequency and $Q$ are programmable, they can be configured as a parallel filter bank or as identical parallel electrode interface channels.
The \ac{PGA} following the \ac{BPF} adds additional gain to the signal but the combined frequency response now takes the response of the \ac{BPF}.

\subsection{\Acl{aADM} encoding}

The basic \ac{ADM} circuit~\cite{Corradi_Indiveri15} belongs to a class of level crossing \acp{ADC} where the sampling interval is adapted based on the characteristics of the coded signal~\cite{inose1966asynchronous}. 
In its classic form it comprises two comparators with a fixed threshold (termed as \emph{delta threshold}) acting as a level shifter around a base line.
This setup outputs an event based on the difference in the amplitude of a signal, when the change is compared with two known thresholds (termed as up- and down- delta thresholds) set by an on-chip \ac{VDAC} to encode both upward and downward swings of the input.
An \emph{`UP'} or \emph{`DN'} tag is attached to the output event based on the polarity of the change, \emph{i.e.},~if the signal has increased or decreased at the level crossing more than the set threshold. The on-chip \ac{AER} interface produces a \emph{`REQ'} signal to transmit the address of the sending node off-chip (see Fig.~\ref{fig:architecture}). Once acknowledged off-chip, the chip-level acknowledge signal \emph{`ACK'} is converted back into a per-channel signal (\emph{`CH\_ACK'}). This signal is gated with a current-controlled ($I_{\mathrm{RFR}}$) inversion stage to create a \emph{`RESET'} mechanism with a refractory period in each channel that resets the \ac{aADM} circuit.

\begin{figure}
\centering
\includegraphics[width=\linewidth]{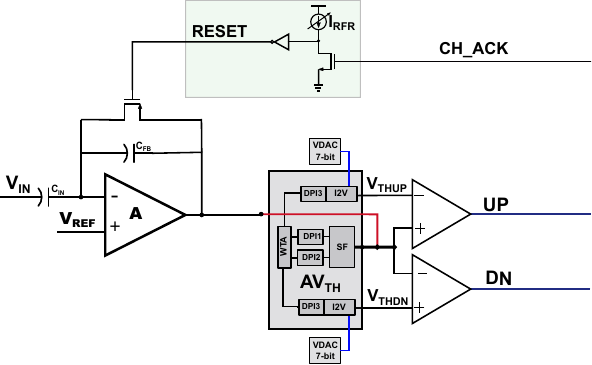}
\Description{Circuit-level schematic of the \ac{aADM} circuit with two comparators and thresholds being generated by the adaptive block, highlighted in grey.}
\caption{Circuit-level schematic of the \ac{aADM} circuit with two comparators and thresholds being generated by the adaptive block, highlighted in grey.}
\label{fig:aadm}
\end{figure}

This approach asynchronously digitizes and encodes relevant changes in the input signal, based on the set delta threshold.
The accuracy of conversion therefore depends on the chosen fixed delta-threshold.
Low thresholds will produce a high data rate, allowing even lossless reconstruction, while higher thresholds will lead to lower data rates, at the cost of lower reconstruction accuracy.
In real-world applications, such as long-term multi-channel monitoring in biomedical or \ac{BCI} domains, input signals are naturally affected by noise, and the noise profile can change and drift with time.
In these cases, a fixed delta threshold might become non-ideal.
The solution implemented here is based on an adaptive scheme that changes the delta threshold based on the low frequency changes of the input signal envelope, but still encoding its high amplitude-high frequency fluctuations~\cite{Sharifshazileh_Indiveri23}.
The schematic diagram of the circuit we designed and fabricated in the \ac{ASIC} presented here is shown in Fig.~\ref{fig:aadm}.

\subsubsection{Adaptive Delta threshold generation}
\begin{figure}
\centering
\includegraphics[width=\linewidth]{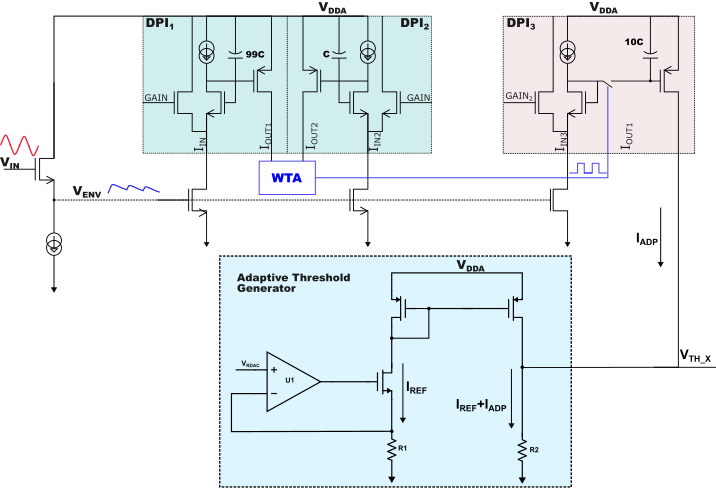}
\Description{Circuit-level schematic of the adaptive delta threshold generator with a source-follower envelope extractor, three \acp{DPI} and one \ac{WTA} block supplying the delta threshold to the comparator}
\caption{Circuit-level schematic of the adaptive delta threshold generator with a source-follower envelope extractor, three \acp{DPI} and one \ac{WTA} block supplying the delta threshold to the comparators of the \ac{aADM} circuit.}
\label{fig:adpvth}
\end{figure}

The adaptive delta threshold generation block consists of four stages.
The input signal from the preceding signal conditioning stage (with gain and filtering circuits) flows into the adaptive delta generator circuit (Fig.~\ref{fig:aadm}, highlighted in grey) with a fixed gain of $A=4$.
Figure~\ref{fig:adpvth} presents the four stages of the delta generator.
The first stage is an envelope extractor which is implemented using a subthreshold \ac{SF} circuit (Fig.~\ref{fig:adpvth}, left).
This extracted envelope is then fed into two \ac{DPI} circuits, which act as current-mode low-pass filters, when operated in the subthreshold regime~\cite{Bartolozzi_etal06,Chicca_etal14}.

The output of the two \acp{DPI} are then compared with a current mode \ac{WTA} circuit to detect a rapid change in signal and to adapt the threshold of the \ac{ADM}.
In the current silicon implementation, the capacitor used in DPI\textsubscript{1} is $99$ times that of DPI\textsubscript{2} to produce a large difference in time constants between the two filters, so that the circuit can determine fast changing signals. The \ac{WTA} picks the winner
and triggers a sampling operation in DPI\textsubscript{3}~\cite{Sharifshazileh_Indiveri23}. The output current of DPI\textsubscript{3} therefore is proportional to the very low-frequency components of the input signal envelope, and neglects the fast transients (such as action potentials)  present in the signal. This current is then added to or subtracted from the baseline threshold, encoded as current and generated by a current to voltage converter (marked as Adaptive Threshold Generator in Fig.~\ref{fig:adpvth}). The composite current is then converted back to voltage within the same block, generating the final adaptive delta threshold voltage (VTH\textsubscript{X}). For the `UP' threshold the adaptive current is added to the baseline, and for `DN' one it is subtracted from the baseline.

\subsection{Pulse frequency modulation}
The \ac{PFM} encoder is realized using an \ac{AdExp} circuit~\cite{Rubino_etal20} using previously proposed topology schemes~\cite{Narayanan_etal23}.
This result is obtained by first converting the amplified and filtered voltage into a current, and then by sending it directly as input to the \ac{AdExp} silicon neuron.
The voltage-to-current conversion is carried out by  a wide-range transconductance amplifier with source degeneration to extend the linearity.
The current is then full-wave rectified and subsequently copied to the neuron input.
As \ac{PFM} encoding of signals has been amply described in the literature and demonstrated across many implementations, in this work we focus mainly on the \ac{aADM} results.
However, preliminary measurements on this chip proved this section to be functional as well.


\section{Measurement results}
\label{sec:measurement-results}
The initial silicon measurements reported in this section verify the chip’s functionality.
\subsection{Noise measurement}
Noise analysis of a typical analog front end can be modeled as: 
\begin{equation}
    v_{n,AFE}^{2} = v_{n,LNA}^{2} + \frac{v_{n,BPF}^{2}}{A_{LNA}^{2}} + \frac{v_{n,PGA}^{2}}{A_{LNA}^{2} A_{BPF}^{2}} + \frac{v_{n,ENCODER}^{2}}{A_{LNA}^{2} A_{BPF}^{2} A_{PGA}^{2}}
    \label{eq:noise}
\end{equation}
The overall input referred noise of the chain is dominated by the first stage in the chain, namely the \ac{LNA} and its gain.
The gain of the following stages helps in reducing the fractional term in Eq.~\ref{eq:noise}.
As the primary objective of this chip was to demonstrate its on-chip adaptive compression capabilities, in the inherent trade-off of area, power and noise we allocated more area to the
\ac{aADM} circuit at the cost of higher noise levels in the \ac{LNA}, which could, however, be further optimized if necessary (at the cost of more area).


\begin{figure}
\centerline{\includegraphics[width=\linewidth]{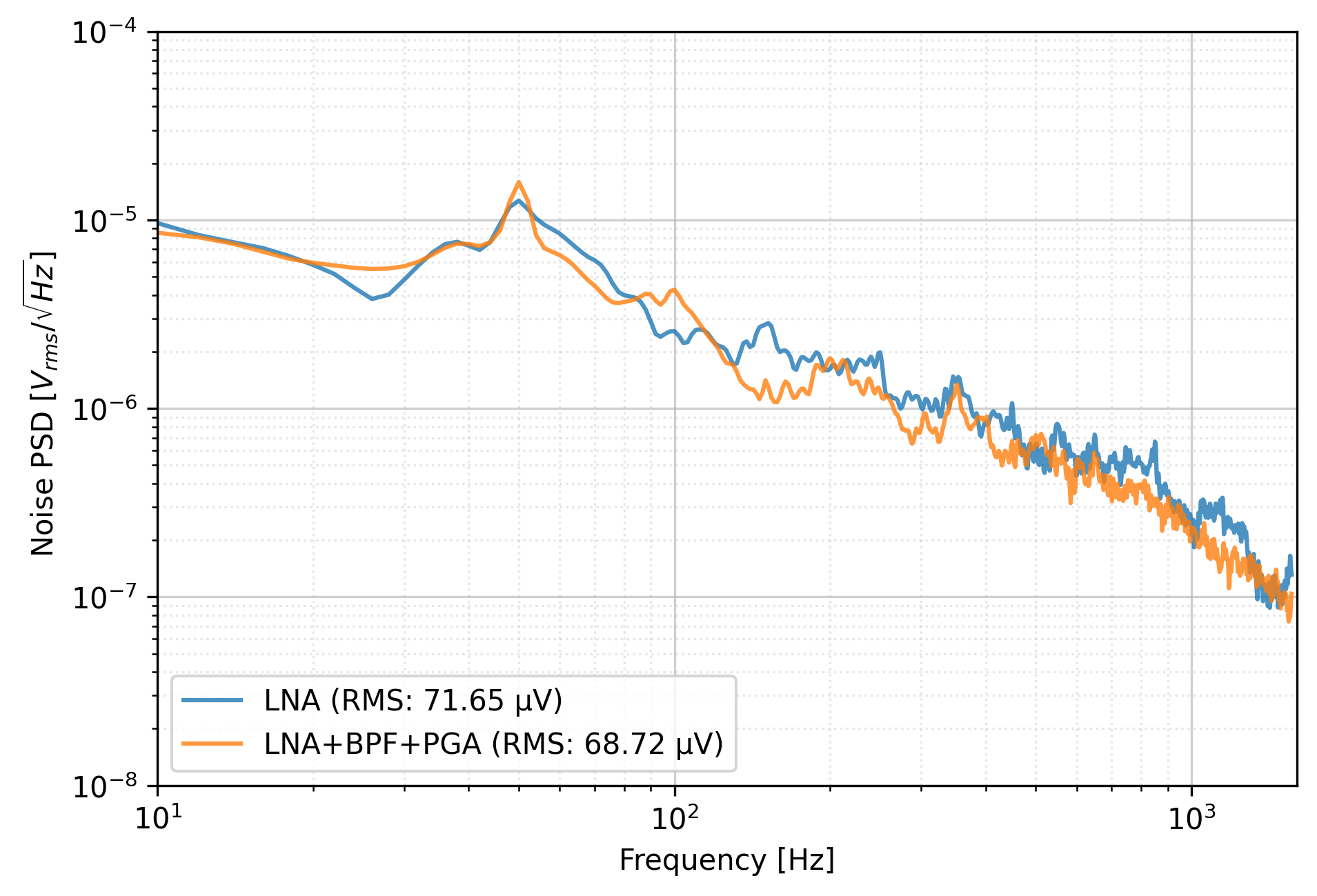}}
\Description{Integrate and Fire Neuron with Full-wave Rectifier}
\caption{Input Referred Noise Power Spectral Density integrated from 10\,Hz to 1.6\,kHz.}
\label{fig:inputreferrednoise}
\end{figure}

\begin{figure}
\centering
\includegraphics[width=0.55\textwidth]{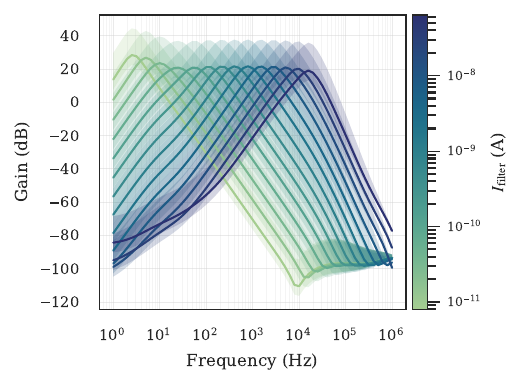}
\Description{Frequency Response of LNA-BPF-PGA Chain and Gain Configurations}
\caption{Frequency Response of LNA-BPF-PGA Chain and Gain Configurations.
The solid lines are the mid-gain code=7, shaded area depicts the 0-15 gain-code range}
\label{fig:pgafreq}
\end{figure}

The initial silicon characterization of the AFE signal chain was performed to quantify the programmable gain control and noise performance. The measurement results of the gain and frequency response of the chain is shown in Fig.~\ref{fig:pgafreq}. The measurement of the LNA-BPF-PGA chain over a target biomedical bandwidth of 1.6\,kHz demonstrated an input-referred noise power spectral density integrating to 68.72\,$\mu\text{V}_{\text{rms}}$.

\subsection{Encoding}

\begin{figure*}
  \centering
  \begin{subfigure}{0.45\textwidth}
    \includegraphics[width=\textwidth]{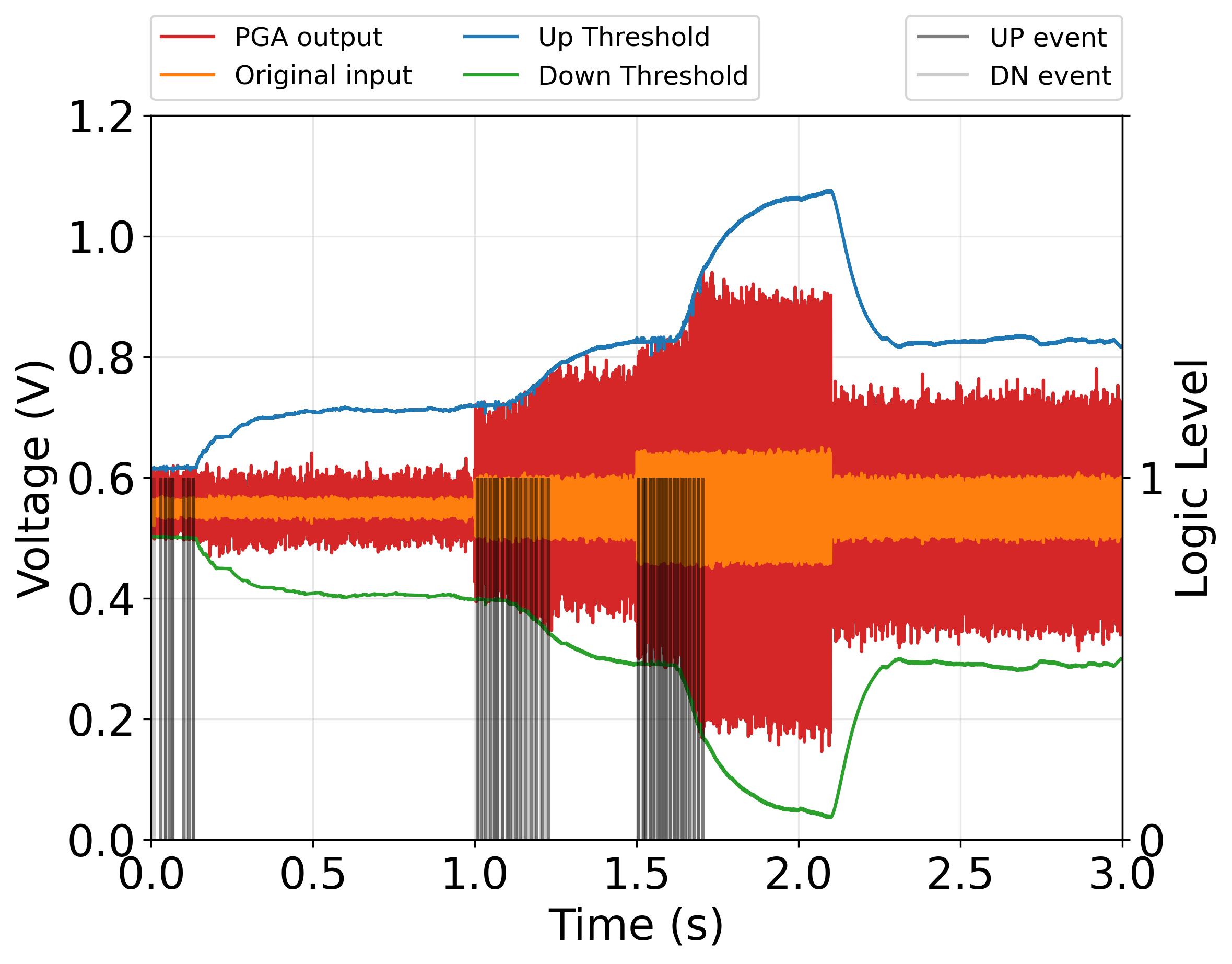}
    \caption{}
    \label{fig:staircasesim}
  \end{subfigure}
  \begin{subfigure}{0.45\textwidth}
    \includegraphics[width=\textwidth]{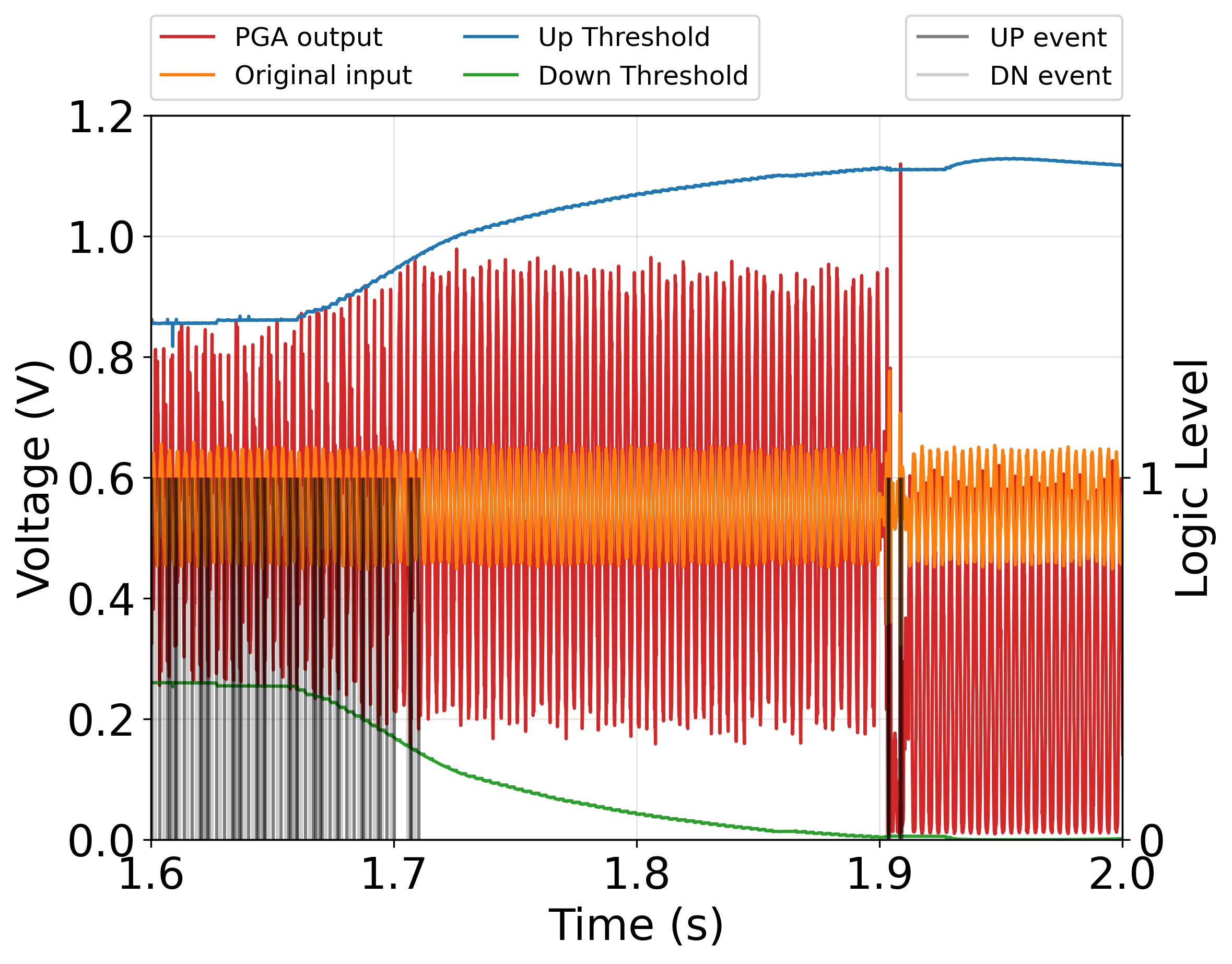}
    \caption{}
    \label{fig:staircasesim-ap}
  \end{subfigure}
  \Description{Circuit-level simulation of the \ac{aADM} block showing changing threshold based on the input signal}
  \caption{Circuit-level simulation of the \ac{aADM} block showing changing threshold based on the input signal; (\subref{fig:staircasesim}) shows a noisy input signal with staircase amplitude, changing threshold, and corresponding encoded events. (\subref{fig:staircasesim-ap}) Zoomed-in part of the simulation, with an additional pulse artificially added to the input signal, to represent an action potential or event of interest at t=1.9s. As expected, the encoder responds to this high-frequency input pulse even after the \ac{aADM} has adapted to the highest noise levels.}
  \label{fig:staircasesimstaircasesim-ap}
\end{figure*}

A circuit-level simulation showing the internal signals of the adaptive delta threshold circuit is shown in Fig.~\ref{fig:staircasesim}. Both an input signal (pure-tone sinusoid), amplitude modulated with a staircase function and added white noise, representing changing noise floors in a realistic control-input are shown. The input was first amplified though the gain stages before being fed into the adaptive generator block. The input to the adaptive generator block as well as the output delta (`UP' and `DN') thresholds  (depicted in blue and green lines) are depicted in relation to the input in Fig.~\ref{fig:staircasesim}. The encoded events are presented as logic pulses showing the fast adaptation with changing noise floors in the staircase signal. Figure~\ref{fig:staircasesim-ap} demonstrates that after adaptation, the addition of an event of interest (e.g., an action potential in a neural recording) will encode it reliably with additional events, despite having adapted to the high noise floor.

\begin{figure}
  \centering
  \includegraphics[width=\linewidth]{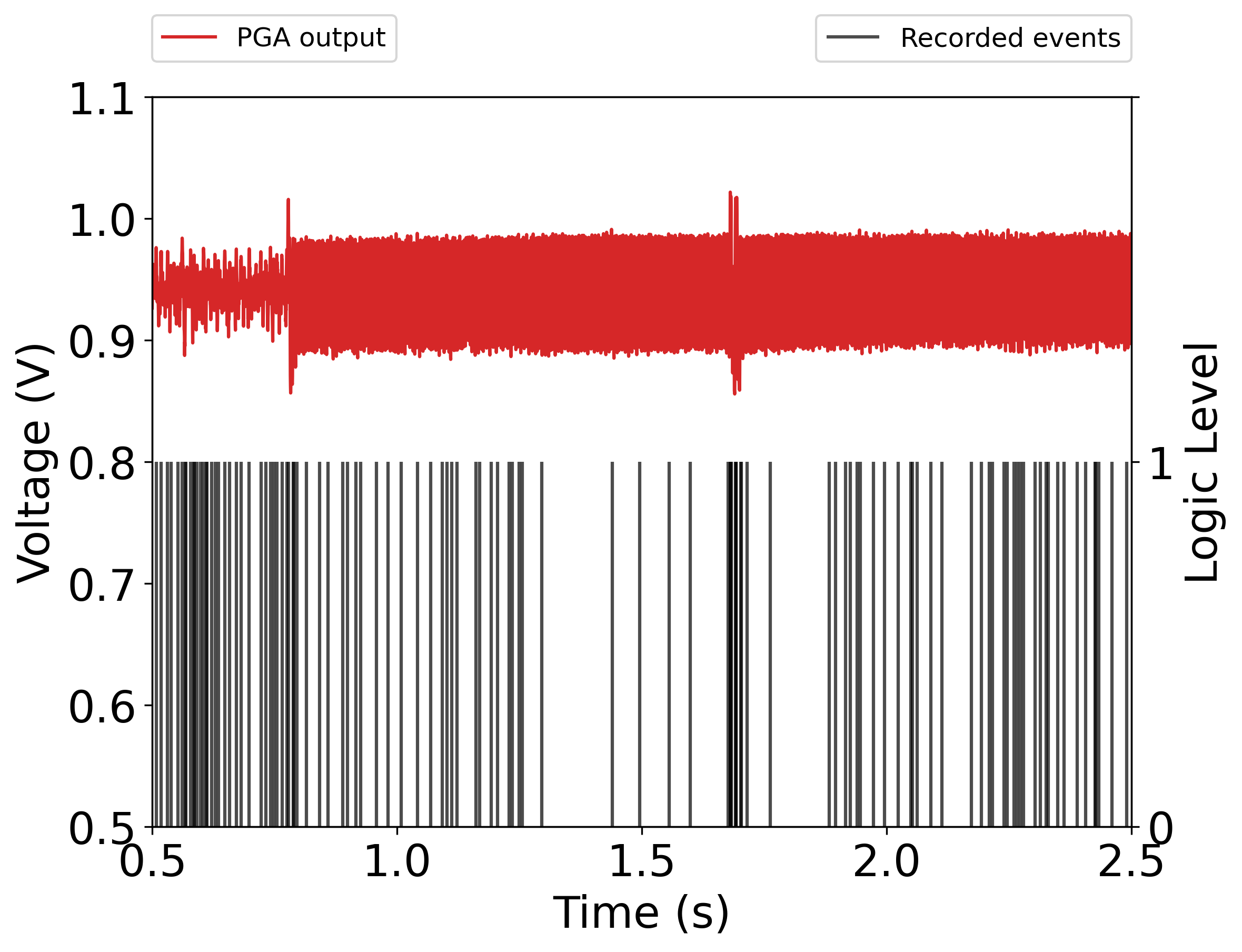}
  \Description{Chip measurements showing \ac{aADM} events in response to the control amplitude modulated step function}
  \caption{Chip measurements showing \ac{aADM} events in response to the control stimulus with step in the noise level at $\mathbf{t=0.6}$\,s approximately, and with an input event of interests at approximately $\mathbf{t=1.7}$\,s. Note that `UP' and `DN' events are merged in the current measurement setup.}
  \label{fig:chipresponse}
\end{figure}
Figure~\ref{fig:chipresponse} shows experimental measurements of an analogous experiment run directly on the ASIC. The synthetic control stimulus was scaled appropriately to $20$\,mV amplitude and played into the chip. In order to match the silicon measurement with simulation, just one channel was enabled with the rest of the channels disabled. The voltage shown in Fig.~\ref{fig:chipresponse} is a real oscilloscope capture of the \ac{PGA} output signal. The events of the \ac{aADM} were recorded at the output in form of the \ac{AER} with external handshaking hardware in the loop. The chip recorded data demonstrates adaptation in the recorded events while encoding the event of interest near $t=1.7$\,s, validating the hardware implementation of the adaptive event-encoding circuit.

\section{Conclusions}
We successfully demonstrated the operation of an event-based analog front-end \ac{ASIC} with 32 parallel channels, featuring extensive programmability via SPI, and supporting the AER output protocol commonly used by neuromorphic \ac{SNN} processors. By implementing an \ac{aADM} encoder alongside a standard \ac{PFM} encoder, the ASIC resolves the critical issue of data saturation in long-term sensor monitoring~\cite{Sharifshazileh_etal21} while providing  sensor-level tunable and adaptive data compression features for a wide variety of Brain-Computer Interface applications. This architectural breakthrough positions the ASIC as a versatile companion front-end for interfacing low-frequency biomedical signals directly with asynchronous \ac{SNN} processors.

\section*{Acknowledgments}
We would like to thank Chenxi Wen the Institute of Neuroinformatics, University of Zurich and ETH Zurich
for carrying out the circuit-level simulations of the \ac{aADM} used in Fig.~\ref{fig:staircasesimstaircasesim-ap}.
Part of this work was supported by the Swiss National Science Foundation (SNSF projects 204651 and 217160).
During the preparation of this work, the author(s) used generative AI tools to assist with language editing and revising the manuscript text, including improving sentence structure, clarity, and adherence to academic style conventions. The tool(s) was used solely to improve clarity, grammar, and readability of author-written content. The tool(s) was not used to generate original scientific content, data, analysis, or conclusions. After using this tool, the author(s) reviewed and edited the content as necessary and take(s) full responsibility for the accuracy and integrity of the final content of the article.

\bibliographystyle{ACM-Reference-Format}
\bibliography{biblioncs,refs}


\end{document}